\documentclass{article}
\usepackage{spconf,amsmath,graphicx}

\usepackage{enumitem}
\setlist{nosep, leftmargin=14pt}

\usepackage{dirtytalk}
\usepackage{booktabs}
\usepackage{multirow}
\usepackage{svg}
\usepackage{microtype}
\usepackage{subcaption}
\usepackage{rotating}
\usepackage{microtype}
\usepackage{caption}
\usepackage{hyperref}

\title{Interpretable 2D Vision Models for 3D Medical Images}
\name{Alexander Ziller$^1$
\qquad Ayhan Can Erdur$^1$ 
\qquad Marwa Trigui$^1$
\qquad Alp G\"uvenir$^1$
\qquad Tamara T Mueller$^1$ }{
\qquad Philip M\"uller$^1$ 
\qquad Friederike Jungmann$^1$ 
\qquad Johannes Brandt$^1$ 
\qquad Jan Peeken$^1$ }{
\qquad Rickmer Braren$^1$ 
\qquad Daniel Rueckert$^{1,2}$
\qquad Georgios Kaissis$^{1,2,3}$
}

\address{$^1$Technical University of Munich, $^2$ Imperial College London, $^3$Helmholtz München
}
\begin{document}
%
\maketitle
\begin{abstract}
Training Artificial Intelligence (AI) models on 3D images presents unique challenges compared to the 2D case: Firstly, the demand for computational resources is significantly higher, and secondly, the availability of large datasets for pre-training is often limited, impeding training success. 
This study proposes a simple approach of adapting 2D networks with an intermediate feature representation for processing 3D images. 
Our method employs attention pooling to learn to assign each slice an importance weight and, by that, obtain a weighted average of all 2D slices.
These weights directly quantify the contribution of each slice to the contribution and thus make the model prediction inspectable. 
We show on all 3D MedMNIST datasets as benchmark and two real-world datasets consisting of several hundred high-resolution CT or MRI scans that our approach performs on par with existing methods. 
Furthermore, we compare the in-built interpretability of our approach to HiResCam, a state-of-the-art retrospective interpretability approach.
\end{abstract}
%

\section{Introduction}
Most computer vision applications focus on 2D image analysis, such as photographs. Medical imaging, in contrast, stands out by capturing primarily 3D images through techniques like computed tomography (CT) or magnetic resonance imaging (MRI). Adapting existing 2D models to 3D data is not straightforward, as it poses challenges in obtaining high-quality pre-trained weights from large datasets \cite{yang2021reinventing}. Additionally, training 3D models requires significantly more computational and memory resources compared to their 2D counterparts \cite{zhang2021liver}. 
In response to these considerations, researchers have explored 2.5D architectures as a compromise, aiming to combine the benefits of strong pre-training with reasonable computational requirements while leveraging the inherent nature of 3D volumes.
These 2.5D architectures often rely on recurrent neural networks (RNN)~\cite{qiu2021recurrent}, e.g. long and short term memory units (LSTM)~\cite{hochreiter1997long}.

Despite their strong performance, deep learning models are often criticised as \say{black-boxes}, where only input and output are observed, but the internal connections remain obscure. To address this concern, interpretability methods have been developed, most prominently Grad-CAM and its successors, which allow for the retrospective inspection of these connections \cite{selvaraju2017grad,draelos2020use}.
Another line of work in this direction was the introduction of attention mechanisms. 
Initially developed for natural language processing, attention mechanisms have also demonstrated impressive performance in vision tasks \cite{dosovitskiy2020image}. These approaches are based on the intuition of \say{learning what is important} and thus prioritising the influence of \say{more relevant} input on the model's prediction. Attention mechanisms have also been extended to improve pooling operations \cite{radford2021learning}, which are commonly applied as a maximum or averaging operation over large inputs in neural networks.

\begin{figure}[t!]
    \centering
         \includegraphics[width=.42\textwidth]{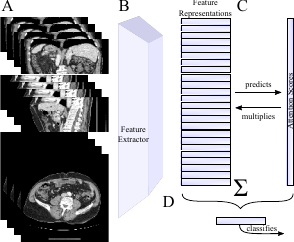}
    \caption{Overview of our approach. (A) The 3D image gets disassembled into slices along all axes. (B) Using the feature extractor from a 2D model each slice obtains a feature representation. (C) Based on the features an attention map is predicted, which assigns importance scores, which sum up to one to the slices. The slice features are multiplied by their attention score. (D) The weighted slice features are summed into a single feature vector. This is used as input for the classifier. Example image stems from the MSD Liver dataset \cite{antonelli2022medical}}
    \label{fig:methods}
\end{figure}
In this study, we propose a conversion strategy for arbitrary 2D models with intermediate feature representations for 2.5D architectures. 
Our method offers several distinct advantages.
Firstly, it is lightweight and only introduces a minimal increase in parameters. Secondly, our approach allows for the utilisation of strong pre-trained weights that are available for 2D architectures. 
Thirdly, we implicitly generate an attention map during the forward pass, which allows for the attribution of individual slices to the final prediction. 
Opposed to retrospective methods, such as Grad-CAM \cite{selvaraju2017grad}, we obtain this attribution \say{for free} during the forward pass.
Furthermore, we obtain quantifiable importance measures, not indirectly derived by inspecting gradients, but directly used for the prediction.  
This leads to an inherent interpretability of our method.
Our experiments demonstrate the efficacy of our method in both, model performance and prediction interpretability. Our code is available at {\footnotesize\url{https://github.com/TUM-AIMED/2.5DAttention}}.

There exist several lines of work, which utilise 2.5D architecture designs. We discuss here the two most relevant. 
Yang et al. \cite{yang2021reinventing} developed methods to make pre-trained 2D architectures applicable to 3D images. They apply 2D convolutional filters along all three axes and thus create an artificial 3D kernel called ACS-convolution. Similar to our approach, this allows them to re-use information learned in the 2D domain while keeping the number of parameters almost constant. 
The work by Li et al. \cite{li2022satr} combines 2D neural networks as feature extractors with transformer blocks for lesion detection. This allows them to learn the attention over all slices of the 3D input and combine them into a single representation. 
There are numerous methods to quantify inputs' contribution to a model's output. Most prominently, we note Grad-CAM \cite{selvaraju2017grad} and its successors, such as HiResCam \cite{draelos2020use}. However, most of them are based on an indirect retrospective inspection of a trained model. Using attention pooling yields such a contribution quantification directly during the forward pass, making predictions inherently interpretable. 

In this work, we introduce an intuitive and versatile approach for converting 2D to 2.5D models for 3D image analyses:
\begin{itemize}
    \item We propose a framework that adapts a 2D network for 3D tasks through slice-wise feature extraction and attention-pooling aggregation.
    \item We demonstrate the effectiveness of our approach on several benchmark datasets as well as two real-world datasets and show that it performs on par with previous methods.
    \item We demonstrate the inherent interpretability of our method and compare it to HiResCam as a retrospective state-of-the-art interpretability method.
\end{itemize}
By this, we hope to contribute towards high-performing and interpretable AI for medical image analysis.

\section{Methods}
We leverage the prevalent structure commonly found in deep learning architectures, which consists of a feature extraction module followed by a classifier. Building on this pattern, we decompose the 3D image into 2D slices along all axes. These individual slices are then fed into the feature extraction component of the 2D network, resulting in the retrieval of $H + W + D$ feature representations, where $H$ represents the height, $W$ the width, and $D$ the depth of the 3D image. To consolidate these feature representations into a single comprehensive unit, we employ attention pooling.

Attention pooling utilises a Multihead Attention layer \cite{vaswani2017attention}, which takes query, key, and value inputs. Following the suggestion by Radford et al. \cite{radford2021learning}, we employ the mean of all slice features as the query, while the key and value comprise the complete set of features. By doing so, our network is capable of estimating the importance of each feature and generating a weighted combination of the input features that best represents the input scan for a given task. This approach generates attention maps during the forward pass, providing insights into the contribution of each individual slice to the final feature representation.
The resultant aggregated feature representation is then fed into the classification layer for prediction. For a visualization of this process compare also Figure \ref{fig:methods}.

By adopting this framework, we aim to merge the inherent advantages of both 2D and 3D architectures, leveraging the strong pre-training available for 2D models while processing the complete 3D volumes. The attention pooling mechanism not only facilitates feature reduction but also implicitly generates attributable predictions through the attention maps. These attention maps quantify the contribution of individual slices to the overall feature representation, aiding in the comprehension of the decision-making process of our model.

\section{Results}
\subsection{Datasets}
We use all 3D datasets from MedMNIST \cite{medmnistv2}, which stem from real-world medical datasets that have been pre-processed and standardised to a resolution of $28\times 28\times 28$ to be used as benchmark problems. They represent diverse medical AI problems and present distinct challenges. 
For more details, we refer to \cite{medmnistv2}.

\begin{table*}[h!]
    \centering
    \begin{tabular}{@{}lllllllllllll@{}}\toprule
        &\multirow{2}{*}{Method}&& \multicolumn{6}{c}{MedMNIST 3D} &\multirow{2}{*}{PDAC}&\multirow{2}{*}{Glioma} \\
        &&& Organ & Nodule & Fracture & Adrenal & Vessel & Synapse \\ \midrule
        \multirow{4}{*}{\cite{yang2021reinventing}}&\multirow{2}{*}{3D-Conv}  & AUROC &  96.5\% & 58.1\% & 57.9\% & 80.7\% & \textbf{92.3\%} & 61.9\% &99.8\%&90.5\% \\ 
        && ACC &77.8\% & 74.0\% & 43.4\% & 79.9\% & \textbf{93.8\%} & 66.1\% & 98.6\% &85.1\% \\ \cmidrule{2-11}
        &\multirow{2}{*}{ACS-Conv} & AUROC &  99.2\% & 79.0\% & 58.1\% & 79.1\% & 84.3\% & 61.5\% & \textbf{100\%} & 93.2\%\\ 
        && ACC & 89.7\% & 81.9\% & \textbf{43.8\%} & \textbf{80.1\%} & 91.0\% & 71.1\% & \textbf{99.8\%}&\textbf{87.9\%} \\ \midrule
        \multirow{2}{*}{Ours}&\multirow{2}{*}{AttentionPool} & AUROC & \textbf{99.5\%} & \textbf{82.3\%} & \textbf{58.8\%} & \textbf{80.9\%} & 91.4\% & \textbf{74.3\%} & 99.1\%&\textbf{93.3\%} \\
        && ACC & \textbf{92.0\%} & \textbf{83.9\%} & 42.7\% & 78.9\% & 91.8\% & \textbf{75.5\%} & 98.6\%& 86.8\%\\\bottomrule
    \end{tabular}
    \caption{\textbf{Results of all approaches on various datasets}. AUROC: area under the receiver-operator curve, ACC: classification accuracy. 
    All experiments were performed five times and the mean result is reported. }
    \label{tab:results}
\end{table*}

\begin{figure}[t!]
    \centering
    \includegraphics[width=0.49\textwidth]{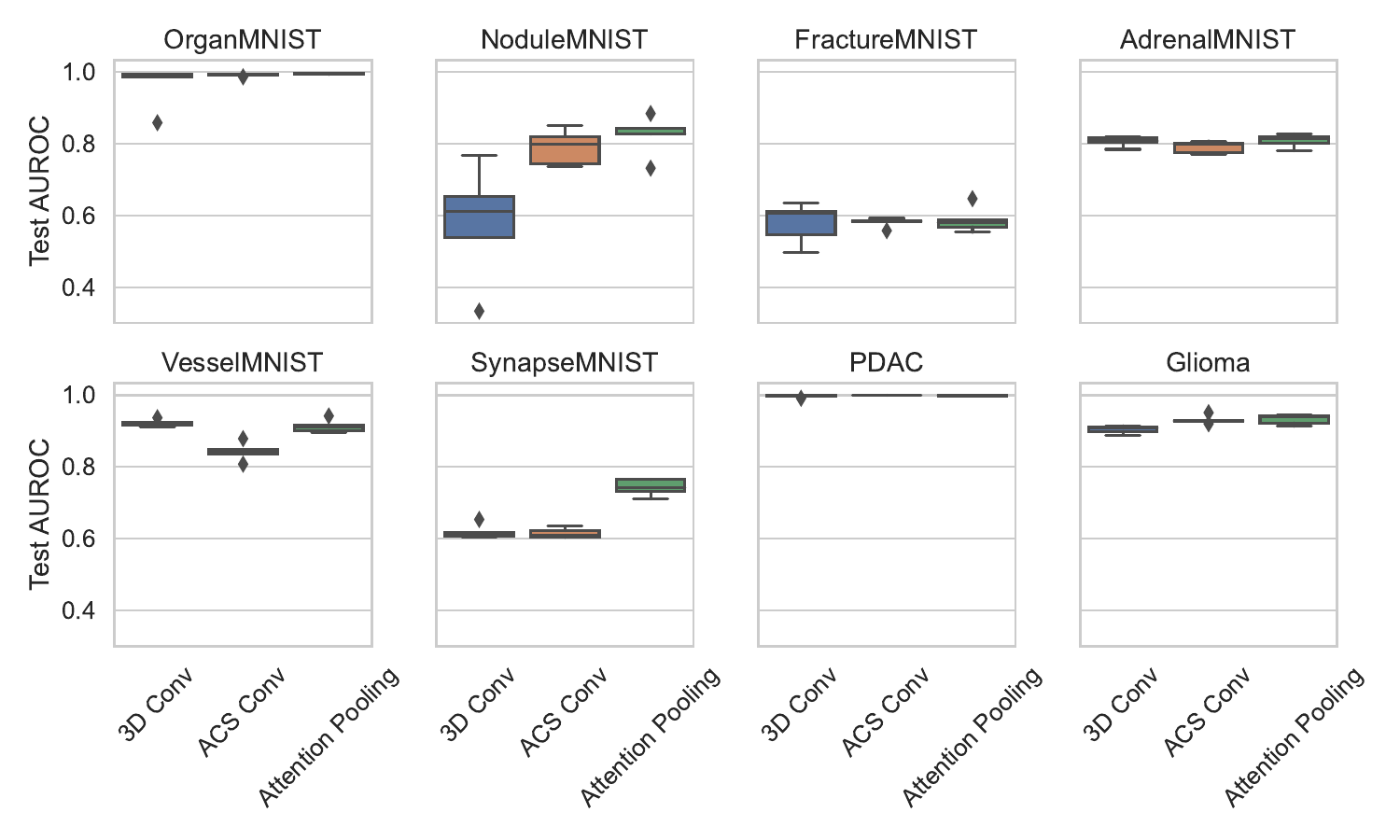}
    \caption{Boxplots of the AUROC for the different MedMNIST3D datasets and the different approaches. }
    \label{fig:mnist_results}
\end{figure}
To showcase the clinical applicability of our approach, we incorporate an internal dataset of 1625 full-resolution abdominal CT scans, classifying the presence of pancreatic ductal adenocarcinoma (PDAC). Out of the total dataset, 867 patients are diagnosed with a tumour by a panel of medical experts, while the remaining 758 CT scans are obtained from patients with diagnoses unrelated to the pancreas, serving as the control group. All images encompass the relevant body regions and are acquired following the same imaging protocol. We split this dataset into 975 CTs as train images and 325 for validation and testing each. We ensure that the distribution of classes in each subset is approximately equal. Additionally, all CT scans are resized to a standardised resolution of $224\times 224 \times 128$ voxels and clipped to an abdominal window of $(-150, 250)$ Hounsfield units. \\
As a second clinically relevant real-world dataset, we classify glioma tumour subtypes from preoperative brain MRI scans. We use three publicly available datasets: UCSF \cite{calabrese2022university}, EGD \cite{van2021erasmus} and TCGA \cite{bakas2017advancing}, of which the first two are taken for training, and the latter is held out as an external test set. Of the total of 1173 patients, 849 are diagnosed with glioblastoma multiforme (GBM), and the rest have other subtypes, boiling down the task into an unbalanced binary prediction. We split the training cohort into 767 train and 192 validation patients, maintaining the class distribution. The test cohort contains 214 patients with 139 GBMs. For each patient, we use three 3D MRI scans, i.e., T1-weighted, T1-contrast-enhanced and T2-weighted fluid-attenuated inversion recovery (FLAIR) images. All images are co-registered, resampled to the 1-mm isotropic resolution, and cropped into $96\times 96 \times 96$ voxels around the center of mass of the tumour computed using segmentation labels obtained with BraTS Toolkit \cite{kofler2020brats}. As the images are registered, we stack all modalities into a tensor with multiple channels. 

\begin{table}[th!]
    \centering
    \begin{tabular}{@{}llll@{}}\toprule
        Approach & Params (Mio) & Memory (MiB) & Time (s) \\\midrule
         3D &33.2&  15821& 744.1$\pm$0.21\\ 
         ACS&  11.2&14475 & 222.7$\pm$29.4\\\ 
         Ours&12.2& 19091 &434.3$\pm$0.45\\ 
         \bottomrule
    \end{tabular}
    \caption{Computational requirements on the PDAC dataset. Params: number of trainable model parameters in million. Memory: memory consumption in Mebibyte on a GPU. Time: mean and standard deviation of one epoch in seconds.}
    \label{tab:resources}
\end{table}
\subsection{Setup}
For our experiments, we adopted a ResNet-18 \cite{he2016deep} as the base architecture. Consistent with the approach by Yang et al. \cite{medmnistv2}, we employed the framework by the same authors \cite{yang2021reinventing} to convert the model into 3D and ACS form. We note that in all conversions pre-trained weights from ImageNet are utilised and converted. For the attention pooling, we employed eight heads. 
All our models were trained for 50 epochs, using the NAdam optimiser \cite{dozat2016incorporating}, which almost eliminates the need to tune the learning rate. We used a learning rate of $2\cdot 10^{-4}$ and a batch size of 64. 
During the first two epochs, we exclusively trained the feature reduction and classification modules, which were randomly initialised. This was done to protect the pre-trained weights from large gradients. After two epochs we also trained the feature extractor.
We report the test accuracy (ACC) and area under the receiver-operator curve (AUROC) as evaluation metrics. 

\begin{figure*}[th!]
    \centering
    \includegraphics[width=0.8\textwidth]{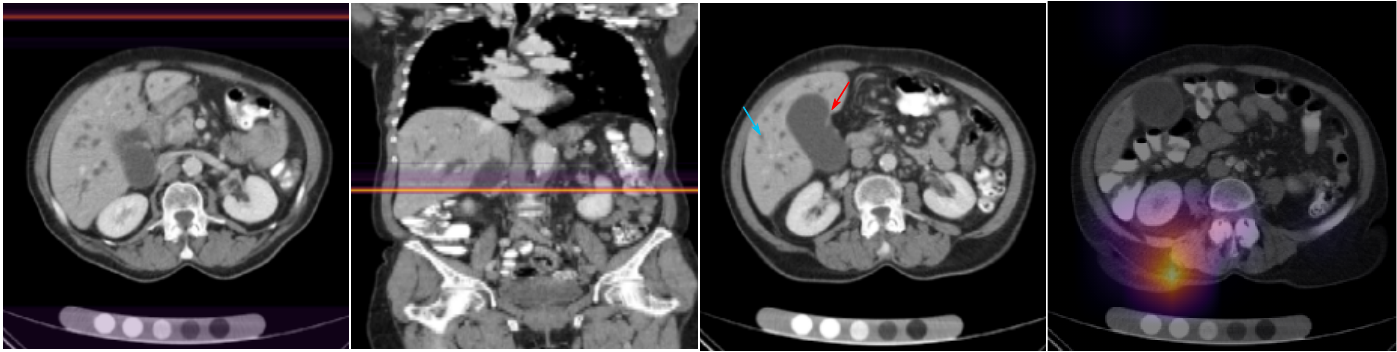}
    \caption{Exemplary image of a correctly classified PDAC tumour from the test set.
    The left panel shows the middle axial slice with overlaid attention values of sagittal and coronal slices. The second panel shows the middle coronal slice with overlaid axial attention values. The third panel shows the slice with the highest attention weight.
    Attention is very sparse and a major part of the slices with non-negligible attention values are on empty parts of the scan. The by far highest attention value is on a slice through the liver, where metastases (blue arrow) and an enlarged gall bladder (red arrow) are visible. Both are strongly indicative of PDAC. The right panel shows the highest peak when applying HiResCam using a trained 3D model to the same CT scan. The highlighted area at the back of the patient contains no features relevant to the task.
    }
    \label{fig:pdac_attention}
\end{figure*}
\subsection{Experiments}
First, we evaluated the same model and weights converted to a 3D, ACS, and our attention-pooling approach on all 3D datasets in MedMNIST. We performed each experiment with five random seeds. We report the results for ACC and AUROC in Table \ref{tab:results} and illustrate the results in Figure \ref{fig:mnist_results}. 
We find that the performance of none of the approaches clearly stands out over all datasets. Our approach has the best results on three of six datasets, the 3D convolution approach has the best on the Vessel task, and the remaining two are ambiguous with AUROC favouring our method and accuracy the ACS convolutions.

To assess the approaches under real-world conditions, we further evaluated them on two high-resolution real-world medical imaging datasets. For the PDAC dataset, we find that all approaches yield almost perfect predictions, yet, the ACS approach stands out with a perfect AUROC score in all five repetitions.  On the glioma dataset, we observe that the ACS approach of \cite{yang2021reinventing} yields the best accuracy while our approach performs best in AUROC. The performance of the 3D ResNet is slightly below the other approaches.  In summary, all approaches perform well on real-world datasets, with ACS and our approach slightly outperforming the 3D model. 
In Table \ref{tab:resources}, we see that while our approach allows for a substantial speed-up compared to a 3D model, ACS is the computationally most efficient method.\\
In addition to performance, attention pooling as a feature reduction module introduces inherent interpretability by quantifying the weight of each individual slice for 3D data classification. Figure \ref{fig:pdac_attention} visualises these capabilities of our architecture design. Upon evaluating several examples on both PDAC and Glioma datasets, we observed that attention is often sparse, with only a small subset of slices receiving the majority of attention weights. Furthermore, we typically have a peak of attention in body regions critical to the task, along with noticeable but not negligible attention on background slices.
One interesting finding is that on the PDAC dataset, despite achieving almost perfect test accuracy, the attention weights were rarely attributed to the tumour or the pancreas, but instead to secondary features.
When comparing our model interpretations to the explanations produced by HiResCam~\cite{draelos2020use} on the 3D model, we see that these highlight the patient's back, where we cannot detect any indication relevant to the task (compare Figure \ref{fig:pdac_attention}).

\section{Discussion}
This work proposes a lightweight and versatile approach combining 2D vision models with attention pooling to create 2.5D architectures. This design facilitates the transfer of information from large pre-trainings on 2D vision datasets to 3D data. We demonstrate its competitive results on several benchmark datasets and two real-world clinical datasets. 
Crucially, our approach provides inherent explanations for predictions, as the impact of each slice on the final prediction is directly quantifiable with the use of attention pooling.

We observed that although they have significantly fewer parameters, 2.5D approaches are performance-wise on par or superior to 3D models. 
This supports our reasoning that 2.5D approaches can show strong benefits over naively expanding 2D CNN architectures to a third dimension, most importantly leveraging pre-trained weights from the 2D domain and computational efficiency. 
By introducing attention pooling, our approach inherently attributes importance to slices. This exemplarily revealed that the PDAC model predominantly focused on secondary features of tumour patients, such as metastases, rather than the tumour or pancreas themselves. Unlike retrospective inspection methods, we get this attribution alongside every model prediction, making it directly available to the user without further computations, substantially lowering the threshold for practical use.

We leave it to future work to explore how to obtain a more fine-grained pixel-level attribution. This could trivially be obtained by multiplying or adding slice attributions along sagittal, coronal, and axial directions. However, such a procedure would not correspond to the intuition of attention and thus it seems preferable to augment our approach with spatial attention mechanisms. 
Furthermore, it would be interesting to combine retrospective inspection methods with our approach to analyse the alignment of attribution. 

We hope to contribute to the development of architectures which are real-world applicable and, at the same time, not exhibit the opacity of \say{black-box} deep learning architectures. We showcase the effectiveness of our approach across various datasets and tasks, providing practitioners with tools to achieve both: competitive performance and interpretable results.

\section{Compliance with ethical standards}
All subjects gave their informed consent for inclusion before they participated in the study. The study was conducted in accordance with the Declaration of Helsinki, and the protocol was approved by the Ethics Committee of Klinikum Rechts der Isar (180/17S).



\section{Acknowledgements}
This work was supported by the European Research Council (Deep4MI - 884622), German Research Foundation, Priority Programme SPP2177, Radiomics: Next Generation of Biomedical Imaging, 
German Cancer Consortium Joint Funding UPGRADE Programme: Subtyping of Pancreatic Cancer based on radiographic and pathological Features, and the Munich Centre for Machine Learning. The funders played no role in the design or execution of the study, nor on the decision to prepare or submit the manuscript.




\small{
\bibliographystyle{IEEEbib}
\bibliography{refs}
}
\appendix
\begin{minipage}{\textwidth}
{\footnotesize
\begin{center}
    \begin{tabular}{llllllllllllll}\toprule
        \multirow{2}{*}{Method}&& \multicolumn{6}{c}{MedMNIST 3D} \\
        & Organ & Nodule & Fracture & Adrenal & Vessel & Synapse \\ \midrule
        \multirow{2}{*}{AveragePool} & AUC & 99.6\% & 86.9\% & 58.3\% & 87.0\% & 93.1\% & 83.7\%\\
        & ACC & 91.4\% & 85.3\% & 43.3\% & 81.9\% & 88.7\% & 80.1\%\\
        \cmidrule{2-8}
        \multirow{2}{*}{MaxPool} & AUC & 99.2\% & 84.9\% & 57.6\% & 76.3\% & 84.3\% & 68.0\% \\
        & ACC &  88.5\% & 84.7\% & 42.3\% & 79.9\% & 89.8\% & 72.2\% \\
        \cmidrule{2-8}
        \multirow{2}{*}{LSTM}  & AUC & 99.4\% & 84.6\% & 56.1\% & 78.9\% & 88.3\% & 70.7\% \\
        & ACC &92.2\% & 82.3\% & 40.2\% & 76.3\% & 89.9\% & 72.0\% \\
        \cmidrule{2-8}
        \multirow{2}{*}{Transformer}  & AUC & 97.1\% & 67.3\% & 58.3\% & 65.9\% & 59.8\% & 62.7\% \\
        & ACC & 78.1\% & 80.8\% & 40.2\% & 76.8\% & 88.7\% & 73.4\% \\
        \cmidrule{2-8}
        \multirow{2}{*}{AttentionPool}  & AUC & 99.5\% & 82.3\% & 58.8\% & 80.9\% & 91.4\% & 74.3\% \\
        & ACC & 92.0\% & 83.9\% & 42.7\% & 78.9\% & 91.8\% & 75.5\% \\
        \bottomrule
    \end{tabular}
    \captionof{table}{Benchmarking of our method using various feature reduction modules on the MedMNIST3D datasets. AUC denotes the area under the receiver-operator curve, ACC the classification accuracy. Mean results over five runs of our experiments. }
    \label{tab:pool_ops}
\end{center}
}
\end{minipage}

\section{Appendix}
Our method is furthermore compatible with other operations to reduce the slice-wise features into a single representation vector. To illustrate this, we have performed experiments with other obvious choices instead of attention pooling. We note that the interpretability capacaties are only given with attention pooling. The results are displayed in Table \ref{tab:pool_ops}.

\end{document}